# Detecting molecules with plasmonic resonators – Analytic expressions and bounds for the sensitivity and figure of merit

*A.Unger and M.Kreiter*

Max Planck-Institut für Polymerforschung, Ackermannweg 10, 55128 Mainz, Germany

unger@mpip-mainz.mpg.de, kreiter@mpip-mainz.mpg.de

Abstract: We derive analytic expressions for the sensitivity and figure of merit for refractive index sensing with plasmonic resonators in the quasistatic limit. Based on these expressions the limits of detection and their origins are investigated. The critical role of limited energy concentration within the resonators surrounding is shown. The classical figure of merit is found to be solely dependent on the ratio of real to imaginary part of the resonators permittivity . Based on these findings we discuss the optimum resonance wavelength and resonator size and shape for single molecule detection.

Refractive index sensing based on localized surface plasmon (LSPR) resonances is to become an important method for the detection of very few or single molecules bound to a surface. It employs small metal particles whose free electron plasma is excited resonantly and creates a tightly confined electrical field near the surface of the particle. The resonance frequency is strongly dependent on the refractive index within the modal volume of the particle. The latter, being 1000 times smaller than the diffraction limit, provides the sensitivity for small dielectric objects down to single molecules. LSPR's have been used to detect changes in bulk refractive index as well as growing dielectric layers on single particles [1-3]. Overviews on their use in biosensing can be found in [4] and [5]. Despite the potential for label free single molecule sensing, no experiments in this direction have been published yet. Single molecule binding events have been detected by labeling molecules with metal particles to increase the refractive index contrast [6].

To measure the performance of plasmonic resonators two quantities are mainly used. The first is the sensitivity s defined as the frequency shift $\Delta\omega$ upon a change in the bulk refractive index around the resonator $\Delta n$.

$$s = \Delta\omega / \Delta n \tag{1}$$

To reach the highest possible sensitivity optimization of particle size and shape has to be done. Many experimental studies have been devoted to this goal, e.g.[7-9]. To compare the sensing capabilities of different resonators, a second measure for the performance, the figure of merit (FOM) was introduced by Sherry et al. given as the sensitivity upon a bulk refractive index change divided by the line width of the resonance $\Gamma$ [9].

$$FOM_{Classic} = s / \Gamma \tag{2}$$

Compared to the wealth of experimental studies how to optimize the sensitivity or the FOM only few results exist on the theoretical side. Miller and Lazarides [10] developed a theory for the sensitivity of plasmonic resonators to bulk refractive index changes, based on a quasistatic approximation and a linearization of the permittivity function. They find that the sensitivity is linearly increasing with the wavelength of the resonance. Otte et. al. speculated from numerical simulations and experiments that the FOM coincides with the ratio of real part to imaginary part of the metals dielectric function [11]. In our previous work [12] we showed that a more general FOM can be derived directly from an expression of the detectability of a change in refractive index from a noisy signal. It contains the classical $FOM_{Classisc}$ by Sherry et. al. as a special case for a specific experimental situation and analyte. Starting from this FOM it is natural to ask for limits of optimization. It is the purpose of this paper to elaborate these limits in a closed form and to investigate their origin. The paper is structured as follows: First the sensitivity in the quasistatic case is derived from first order perturbation theory. Then the limits of the sensitivity are discussed, using known relations [13], yielding remarkably simple expressions. The results are used to derive the general FOM in the quasistatic case. The following discussion reveals that the classical FOM of Sherry et. al. is solely determined by the resonance wavelength of the resonator. The case of single molecule detection is investigated next. Optimum resonance wavelength and resonator shape are derived from the general FOM.

In [12] we derived a perturbative expression for the sensitivity s of a plasmonic resonator:

$$s = \frac{\Delta\omega}{\Delta n} = -\frac{\omega}{n} C \qquad (3)$$

Here the confinement factor C is given by

$$C = \frac{\int_{V_A} \varepsilon \vec{E}^2 dV}{\int_{AllSpace} \varepsilon \vec{E}^2 dV} \qquad (4)$$

Where $\vec{E}$ is the electrical field, $\varepsilon$ the permittivity as a function of position and the upper integral is carried out over the volume of the refractive index change while the integral in the denominator is carried out over all space. In the nondissipative, nondispersive and nonradiating case, C can be interpreted as the energy confinement to the analyte volume. This gives an intuitive view on the sensitivity: the relative frequency shift equals the relative refractive index contrast times the fraction of the total energy contained within the analyte volume. The assumption of a nondispersive medium however does not hold in the visible. Here the frequency dependence of the permittivity of the resonator has to be taken into account in addition to the change induced by an external refractive index perturbation. This is especially important for quasi-static resonances, where the resonance condition is intimately connected to the resonators permittivity (see [14] denominator of equation 5.32):

$$3\varepsilon_A + L(\varepsilon_M - \varepsilon_A) = 0 \qquad (5)$$

Here $\varepsilon_A$ is the permittivity of the ambient $\varepsilon_M$ the permittivity of the metal and L is a shape parameter which is 1/3 for a sphere. This case can be covered by perturbation theory as well. To do so, a permittivity perturbation which accounts for both the change of the permittivity upon binding of an analyte and the change of permittivity of the resonator upon frequency changes is used:

$$\Delta\tilde{\varepsilon} = \Delta\varepsilon + \frac{\partial\varepsilon}{\partial\omega}\Delta\omega \qquad (6)$$

Some algebraic manipulation then leads to the expression for the sensitivity

$$s = \frac{\Delta\omega}{\Delta n} = -2\frac{\omega}{n}\frac{\int_{V_S}\varepsilon\vec{E}^2 dV}{\int_{AllSpace}\varepsilon\vec{E}^2 dV + \int_{AllSpace}\frac{\partial\omega\varepsilon}{\partial\omega}\vec{E}^2 dV} = -2\frac{\omega}{n}\tilde{C} \qquad (7)$$

with has a similar form as the previous expression eq. (3) but the C in is replaced by $\tilde{C}$ which contains now also the resonators dispersion and an additional factor 2. The case of a nondispersive resonator is contained as a special case in this equation.

Now we consider a quasistatic resonance. In this case it was shown[13] that $\int_{AllSpace}\varepsilon\vec{E}^2 dV = 0$ and eq. (7) simplifies to

$$s_{Static} = \frac{\Delta\tilde{\omega}_{Static}}{\Delta n} = -2\frac{\omega}{n}\frac{\int_{V_S}\varepsilon\vec{E}^2 dV}{\int_{AllSpace}\frac{\partial\omega\varepsilon}{\partial\omega}\vec{E}^2 dV} = -2\frac{\Delta n}{n}C_{static} \qquad (8)$$

Since in the quasistatic approximation all fields have the same phase $\vec{E}^2 = \vec{E}\vec{E}^*$ and $C_{static}$ is an expression for the ratio of energy stored in the analyte volume to energy stored in the entire near field . The expression in the integral in the denominator of (8) is the energy density in a dispersive medium [15]. In contrast to the energy interpretation of eq. (4) the quasi-static frequency shift of a dispersive resonator is a factor of two stronger. This effect can be understood as a positive feedback from the resonators dispersion. When the frequency decreases, the permittivity of the resonator decreases as well and thus the electric field is pushed out of the resonator which leads to an additional shift.

The quasistatic sensitivity $s_{Static}$ from Eq. (8) can now be expressed as the sensitivity for a bulk change in refractive index times a weighting factor f.

$$s_{Static} = s_{Bulk} f(\vec{r}_A, V_A) = -2\frac{\omega}{n} C_{Bulk} f(\vec{r}_A, V_A) \qquad (9)$$

where 0<f<1, and f is a function of analyte position and analyte volume given by

$$f = \frac{\int_{V_a} \varepsilon \vec{E}^2 dV}{\int_{Dielectric} \varepsilon \vec{E}^2 dV} \qquad (10)$$

With this expression two effects can be separated. First the previous results for bulk changes in refractive index are contained for f = 1. Second the effect of different non-bulk analytes is separated and contained only in f. For $s_{Bulk}$ remarkably simple expressions hold and will be derived now.

We denote the energy in the dielectric $U_d$ and the energy in the metal $U_m$. Since $C_{bulk}$ is an expression for the ratio of energy $U_d$ stored in the dielectric to the energy stored in the whole nearfield it can be cast to

$$C_{bulk} = \frac{U_d}{U_m + U_d} = \frac{1}{1 + U_m / U_d} \qquad (11)$$

For their ratio $U_m/U_d$ an analytical result was derived in[13]:

$$U_m / U_d = -\frac{d(\omega\varepsilon')}{d\omega} \Big/ \varepsilon' \qquad (12)$$

where ε' is the real part of the permittivity of the metal. it is known that this ratio is always greater than one [13]. With this result the quasistatic sensitivity can be finally written as

$$s_{Static} = -2\frac{\omega}{n} C_{Bulk} f(\vec{r}_A, V_A) = -2\frac{\omega}{n}\frac{\varepsilon'}{\varepsilon' - \frac{d(\omega\varepsilon')}{d\omega}} f(\vec{r}_A, V_A) \qquad (13)$$

This expression can be compared to the results of Miller and Lazarides [10]. Inserting their linear approximation for the metal permittivity into (13) gives the same result as they obtained (equation 18 in [10]), which shows that their theory is contained in ours as a special case.

Since $U_m/U_d$ is always greater then one, a limit on the bulk sensitivity is found

$$\left|s_{Static}\right| =< \frac{\omega}{n} \qquad (14)$$

Since f is always smaller than one, this is an inherent limit for the sensitivity of a quasistatic resonator and the derivation clearly shows its origin: the sensitivity is a function of the ratio of energy stored in the analyte volume to the energy stored in the metal resonator and this ratio is limited to values smaller than one. Interestingly the same limit applies to nondispersive, e.g. Fabry Perot resonators. Here C can in principle be one if the volume of the whole resonator is accessible for refractive index changes but due to of the absence of dispersion the sensitivity is only half which leads to the same limit.

Now that we have an expression for the sensitivity we can continue to find an expression for the FOM which reflects the true ability to detect a binding event. The general FOM for resonator optimization as derived in our previous work [12] is the ratio of frequency shift induced by a binding event and the uncertainty with which this shift can be measured. It is given by

$$FOM = \frac{s}{\Gamma}\sigma^{\alpha} = \frac{\omega}{\Gamma}C\sigma^{\alpha} = QC\sigma^{\alpha} \qquad (15)$$

It does contain not only the sensitivity. but also an expression how accurate frequency shifts can be measured. From this expression the line width Γ and the cross section σ enter the FOM. The expression ω/Γ is identified as the quality factor Q which is given by [16] . The influence of the scattering or absorption cross section σ depends on the dominating type of noise in the instrument, yielding different values for α in each case $(0 \le \alpha \le 1)$. Q in the quasistatic approximation can be expressed solely in terms of the permittivity of the metal at resonance [13]

$$Q_{Static} = \frac{\omega \partial \varepsilon' / \partial \omega}{2\varepsilon''} \tag{16}$$

Inserting this expression and the expression for $C_{Static}$ , equation (13) into eq (15) the FOM in the quasistatic limit becomes

$$FOM_{Static} = -\frac{\varepsilon'}{\varepsilon''} f(\vec{r}_A, V_A)\sigma^{\alpha} \tag{17}$$

The classical FOM of Sherry et al. which is often used to compare different resonators [9] is obtained by f=1 and α=0. In this case we obtain the simple result

$$FOM_{Classical} = -\frac{\varepsilon'}{\varepsilon''} \tag{18}$$

This is one of the most important results of this paper. It shows that in the quasistatic limit and for bulk changes in refractive index the FOM is already given when the resonance wavelength of the resonator and the material is chosen. It has been argued before based on results from numerical simulations that the ratio of real part to imaginary part of the metal permittivity has the same spectral dependence than the FOM [11]. This is put on a solid ground here and our result shows that it completely determines the FOM in this special case. One could now ask how the FOM behaves when deviations from the quasistatic regime occur. Since the sensing happens in the nearfield and the nearfield is always quasistatic[16] deviations will first appear as additional radiation damping which lowers Q. As a result equation (18) can be regarded as an upper limit for the FOM. This might be different for hybrid modes which are not confined to the metal surface but since these modes are photonic in nature they will be governed by the diffraction limit. As a consequence they might be superior for sensing of changes in bulk refractive index, but not for the sensing of events that happen at the surface of the metal resonator since their spatial confinement to the surface will be poor.

Our derivation shows that for bulk changes in refractive index no further optimisation beside choosing for each metal the optimum resonance frequency is possible. For smaller analytes it is still possible to optimize s and the FOM by changing the size of the resonator relative to the analyte size. It was shown in our previous work, that the scaling and optimization of the FOM with resonator size depends on the type of noise in the optical instrument and the size and shape of the analyte. The scaling is now given by the geometric scaling of f and σ and all previously derived scaling laws are still true. Generally the FOM scales with a linear size of the resonator $r_R$ as

$$FOM(r_R) \propto f(r_R)\sigma(r_R) \propto r^{\beta} \qquad (19)$$

The exponent $\beta$ is given from table 1 in [12].

One of the most interesting sensing cases is the detection of a single molecule binding to the surface of the resonator. It will now be discussed which optimization options remain for this case. We assume that by a suitable light source in the experimental setup the signal can be made large enough to neglect background noise and scattering is measured. The remaining noise sources [12] are then signal noise, i.e. noise associated with the counting statistics of the detection process and instrument noise, i.e. random fluctuations of the resonance spectrum. The latter gives a $\beta = -3$ while the former gives $\beta = 0$. This means that in the signal noise regime the FOM is independent of the resonator size and otherwise it is advantageous to make the resonator as small as possible (provided one never reaches the background noise level). Additionally the confinement factor f can be increased by introducing a sharp tip to the resonator where the field is concentrated.

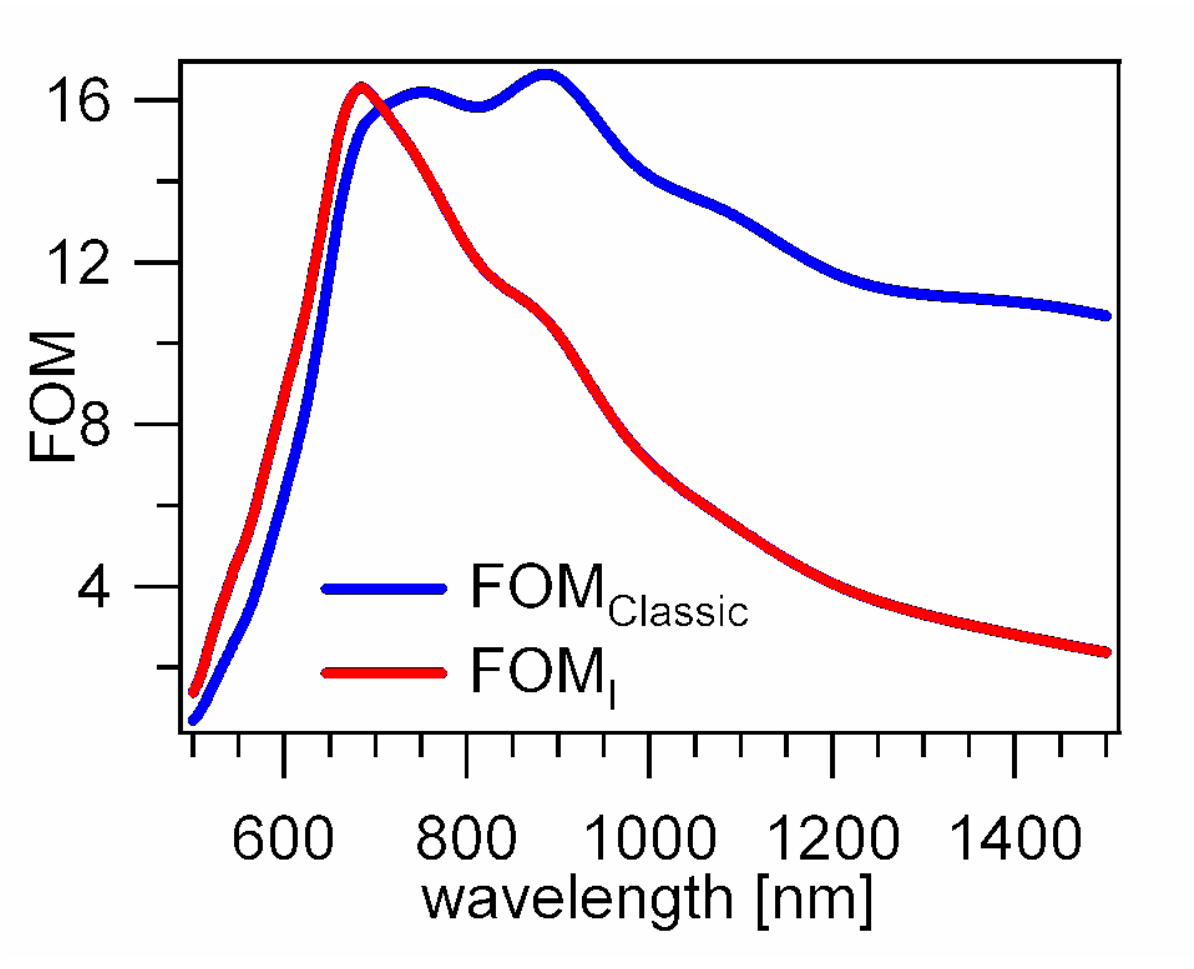

Fig. 1. The classical FOM and the wavelength dependent part of the FOM for signal noise and scattering measurements ($FOM_I$) as a function of wavelength. Since $FOM_I$ depends on the cross section it is scaled to make it comparable with the classical FOM.

As was shown in eq. (17) the wavelength dependence of the FOM is determined by the permittivity of the metal times the scaling of the cross section with wavelength. For instrument noise the cross section does not enter the expression because only the maximum achievable resonance shift matters, and the wavelength scaling reduces to eq. (18) For the signal noise case an additional factor $\lambda^{-2}$ enters the wavelength scaling since the scattering cross section scales with $\lambda^{-4}$. These two functions are shown in Fig. 1 for a gold resonator. A maximum around $\lambda = 700nm$ is observed for both cases, while in the instrument noise regime an additional maximum around $\lambda = 900nm$ is found. This result suggests to choose a resonator with resonance wavelength of $\lambda = 700nm$ and a sharp corner where the field is concentrated. Possible candidates with this properties are gold pyramids with a round base or gold rods with an aspect ratio of roughly 3.5 [14].

In summary we have extended our previously introduced sensing theory to the case of strongly dispersive resonators. Based on this theory analytical expressions for the energy confinement and quality factor of localized plasmons were then used to find simple but general expressions for the sensitivity as well as for the FOM in the quasistatic limit. It turned out that the classical FOM in this limit is solely determined by the ratio of real to imaginary part of the resonators permittivity at the resonance frequency. It therefore is fixed once the resonance frequency is determined. Moreover it was shown that the sensitivity as well as the FOM have upper limits which are caused by a limited quality factor and energy confinement to the sensing volume. The analytical expressions for the FOM were then used to determine the optimum resonance wavelength of a plasmonic resonator which turned out to be around 700 nm for gold. For analytes that do not completely fill the modal volume of the resonance the field confinement to the analyte can still be optimized by optimizing the shape of the resonator. It is suggested that particles with sharp tips, like pyramids should be used to detect single molecules.

## REFERENCES

[1] A. D. McFarland and R. P. Van Duyne, Nano Lett. **3**, 1057 (2003).
[2] J. J. Mock, D. R. Smith, and S. Schultz, Nano Lett. **3**, 485 (2003).
[3] G. Raschke, S. Kowarik, T. Franzl, et al., Nano Lett. **3**, 935 (2003).
[4] J. N. Anker, W. P. Hall, O. Lyandres, et al., Nat. Mater. **7**, 442 (2008).
[5] K. A. Willets and R. P. Van Duyne, Annu. Rev. Phys. Chem. **58**, 267 (2007).
[6] T. Sannomiya, C. Hafner, and J. Voros, Nano Lett. **8**, 3450 (2008).
[7] N. L. Bocchio, A. Unger, M. Alvarez, et al., J. Phys. Chem. C **112**, 14355 (2008).
[8] R. Bukasov and J. S. Shumaker-Parry, Nano Lett. **7**, 1113 (2007).
[9] L. J. Sherry, S. H. Chang, G. C. Schatz, et al., Nano Lett. **5**, 2034 (2005).
[10] M. M. Miller and A. A. Lazarides, J. Phys. Chem. B **109**, 21556 (2005).
[11] M. A. Otte, B. Sepulveda, W. H. Ni, et al., Acs Nano **4**, 349.
[12] A. Unger and M. Kreiter, J. Phys. Chem. C **113**, 12243 (2009).
[13] F. Wang and Y. R. Shen, Phys. Rev. Lett. **97** (2006).
[14] C. F. Bohren and D. R. Huffman, *Absorption and Scattering of Light by Small Particles* (Wiley-VCH, Berlin, 1998).
[15] L. D. Landau and E. M. Lifshitz, *Lehrbuch der theoretischen Physik VIII Elektrodynamik der Kontinua* (Akademie Verlag Berlin, 1985).
[16] J. D. Jackson, 1998), Vol. 2009.